\documentstyle[aasms4,12pt]{article}
\eqsecnum
\def\beq{\begin{equation}}
\def\eeq{\end{equation}}
\def\ref{\reference}
\def\simge{\mathrel{%
   \rlap{\raise 0.511ex \hbox{$>$}}{\lower 0.511ex \hbox{$\sim$}}}}
\def\simle{\mathrel{
   \rlap{\raise 0.511ex \hbox{$<$}}{\lower 0.511ex \hbox{$\sim$}}}}
\begin{document}
\title{ON THE RAPID SPIN-DOWN AND LOW LUMINOSITY PULSED EMISSION FROM
AE AQUARII}
\author{Chul-Sung Choi$^1$ and Insu Yi$^{2,3}$}
\affil{$^1$Korea Astronomy Observatory, 36-1 Hwaam, Yusong, Taejon 305-348:\\
cschoi@hanul.issa.re.kr}
\affil{$^2$School of Physics, Korea Institute for Advanced Study,
207-43 Cheongryangri, Dongdaemun, Seoul 130-012, Korea: iyi@kias.re.kr}
\affil{$^3$Dept. of Physics, Ewha University, Seoul, Korea}
\begin{abstract}

AE Aqr is an unusual close binary system with a very short white dwarf
spin period, a high spin-down rate, a relatively low quiescent luminosity, 
and clear pulse signals. The exact nature of the large spin-down power has 
not been well explained mainly due to the fact that the observed luminosities 
in various energy ranges are much lower than the spin-down power. We consider 
an unconventional picture of AE Aqr in which an accreting white dwarf,
modeled as a magnetic dipole whose axis is misaligned with the spin axis, 
is rapidly spun-down via gravitational radiation emission and therefore the
spin-down power is not directly connected to any observable electromagnetic 
emission. The rapid spin-down is caused by the non-axisymmetric polar mounds 
of accreted material slowly spreading away from the magnetic poles over
the surface of the star. The accretion proceeds at high altitudes toward 
the magnetic poles of the white dwarf while a large fraction of the inflowing 
material is ejected in a propeller-like manner. Based on the observed 
quiescent X-ray and UV emission, the magnetic field strength is estimated as 
$\sim 1\times 10^5 \eta_x^{-1/2}$ G and the mass accretion rate as 
$\sim 1\times 10^{15} \eta_x^{-1}$ g/s where $\eta_x<1$ is the X-ray radiative 
efficiency. A large fraction of the accreted mass is flung out by the 
propeller action and $\sim 50\%$ of the accreted material arrives at the 
magnetic poles. The electromagnetic dipole emission is expected at the level 
of $\sim 1\times 10^{29}\eta_x^{-1}$ erg/s which suggests that for 
$\eta_x\sim 0.1$, the observed radio luminosity could be well accounted for
by dipole radiation.

\end{abstract}


\section{INTRODUCTION}

AE Aqr, which has a spin period of $P_*=33.08$ s and an orbital period of
9.88 hr (Patterson 1979; Welsh, Horne, \& Gomer 1995), is usually 
classified as a DQ Her-type magnetic cataclysmic variable (CV) or 
an intermediate polar.
It consists of a magnetic white dwarf and a companion star with 
a spectral type of K3 -- K5.
This companion fills its Roche lobe and transfers matter to the
white dwarf (Casares et al. 1996).
However, optical observations (e.g., single-peaked Balmer emission  
lines) indicate that there is little evidence of a Keplerian 
accretion disk in the binary system (Welsh, Horne, \& Gomer 1998 
and references therein).

Pulsations at the spin period are clearly seen from optical to X-ray, 
but there are no radio pulsations at this period (Bastian,
Beasley, \& Bookbinder 1996). 
The optical \& UV pulse profiles show a sinusoidal double peak,   
where the two peaks are separated by 0.5 in phase and their 
amplitudes are unequal (Eracleous et al. 1994). 
On the other hand, the X-ray pulse profile has a sinusoidal single 
peak which suggests that the X-rays have a different origin (Eracleous,
Patterson, \& Halpern 1991; Choi, Dotani, \& Agrawal 1999).
Flares are aperiodic and last for $\sim 10$ min to $\sim 1$ hr. The flares
have been observed for AE Aqr in various wave bands (Patterson 1979;
Bastian, Dulk, \& Chanmugam 1988; Eracleous \& Horne 1996; Choi, Dotani,
\& Agrawal 1999).
During a large flare, the UV \& X-ray luminosities increase by a factor of
3 ($\sim 1 \times 10^{32}$ erg/s in UV and $\sim 2 \times 10^{31}$ erg/s in
X-ray) compared with the quiescent luminosities (the quiescent
luminosity in UV is $\sim 4 \times 10^{31}$ erg/s and $7 \times 10^{30}$
erg/s in X-rays).

De Jager et al. (1994) found that AE Aqr is steadily spinning down
at a rate of $\dot{P_*} = 5.64 \times 10^{-14}$ s/s.
One interesting result from this observation is that the spin-down power,
$L_{sd} = I \Omega_* \dot{\Omega}_* \approx 5 \times 10^{34} 
M_{*,1}R_{*,9}^2$ erg/s where $R_{*,9}$ is the white dwarf radius
in units of $10^9$ cm, $M_{*,1}$ is the stellar mass in units of the solar
mass ${\rm M_{\odot}}$, and $\Omega_*=2\pi/P_*$, is much 
greater than the observed quiescent UV and X-ray luminosities, or even 
the bolometric luminosity.
This fact implies that the spin-down power should be mostly converted 
into different types of energy emission. Eracleous \& Horne (1996) and 
Wynn et al. (1997) proposed a magnetic propeller model in which most of the 
accreted matter is expelled from the binary system. 
In this picture the spin-down power is consumed to expel the accreted 
matter. 

Although the propeller model offers a good explanation for the 
observational features of AE Aqr, there is no direct evidence of
the high-velocity gas stream escaping from the system. 
In addition, the propeller model requires a high mass-transfer rate,
$\dot{M} \gtrsim 7 \times 10^{18}$ g/s (e.g. Ikhsanov 1998) in order to 
account for the observed spin-down rate.
This mass-transfer rate is greater by one order of magnitude or 
more than the rate expected from the empirical $\dot{M}$--period
relation (Patterson 1984).
Alternatively, several studies have invoked the pulsar-like spin-down 
mechanism, where the spin-down power can be used for the
generation of electromagnetic dipole radiation and particle acceleration
(e.g., de Jager 1994, Ikhsanov 1998).
According to the recent study by Ikhsanov (1998), the rapid spin-down rate
of AE Aqr can be explained by this mechanism if the white dwarf has
a strong surface magnetic field of $\sim$50 MG.
This field strength exceeds the upper limit of $\sim 5$ MG derived
by Stockman et al. (1992). There has been some sporadic attempts to
connect the observed spin-down power to the yet-to-be-confirmed TeV $\gamma-$ray
emission of $\sim 10^{32}$ erg/s (Meintjes et al. 1992, 1994; Bowden
et al. 1992), which is not convincing. The exact nature of the
spin-down power in AE Aqr therefore remains yet to be clarified.

In this paper, we propose that the gravitational radiation emission 
mainly drives the spin-down power while accretion, ejection, and 
electromagnetic radiation from spinning white dwarf are responsible for
the observed luminosities in various energy bands. 
We construct a self-consistent picture for AE Aqr by removing the large 
spin-down power from the observable electromagnetic emission.
For our numerical estimates, we adopt $R_*= 7\times 10^8$ cm,
$M_*=0.8 \ {\rm M_{\odot}}$, and the moment of inertia $I_*= 3 \times 10^{50}$
g cm$^2$.

\section{SPIN-UP AND SPIN-DOWN IN AE AQR}

AE Aqr contains a very rapidly spinning white dwarf with its spin period
$P_*=33.08$ s. It is an unusual white dwarf as its spin period is quite
close to the theoretically maximum break-up spin period. Such a short
spin period could be achieved through accretion only if the accreted mass
$\Delta M$ over time $\Delta t$ is at least as high as
\beq
\Delta M\simge I_*\Omega_*/(GM_*R_*)^{1/2}\sim 0.1 \ {\rm M_{\odot}},
\eeq
where this estimate has to be taken as a lower bound since we have assumed
that the accreted material has the specific angular momentum $(GM_*R_*)^{1/2}$
which is realized only when the Keplerian accretion disk extends all the way 
down to the
stellar surface. In reality, magnetic truncation could limit the specific
angular momentum to a much lower value than the Keplerian value at the stellar 
surface (e.g. Frank, King, \& Raine 1992; see below).
For an accretion rate ${\dot M}=10^{16} {\dot M}_{16}$ g/s, the accretion
of angular momentum has to occur for the duration of
\beq
\Delta t\simge 6\times 10^7(\Delta M/0.1 M_{\odot}){\dot M}_{16}^{-1} 
\ {\rm yr}.
\eeq
Although such an accretion is possible, it is unclear how the accretion flow
-- magnetosphere interaction could have affected the spin-up of AE Aqr.
Presently, there is no indication of well-defined rotation of accreted material
in the form of the accretion disk. If there is no accretion disk in AE Aqr,
accretion could occur via the diskless accretion such as a ballistic accretion 
stream. The estimated spin-up time scale is short enough to be realized in
binary systems similar to AE Aqr where the secondary is a K3 -- K5 main sequence
red dwarf.

There have been discussions on possible spin-down mechanisms without a clear
favorite. The main problem for various spin-down mechanisms arises due to
the fact that the inferred large spin-down power is not observed in any 
detectable forms such as high velocity gas or high luminosities. 
The relatively low quiescent
luminosities also pose a serious problem for any mechanisms involving 
accretion. If the white dwarf's dipole-type magnetic field is strong enough,
the electromagnetic power due to dipole radiation could account for the
rapid spin-down as discussed by Ikhsanov (1998). The electromagnetic power
is estimated as
\beq
L_{em}=2\mu_*^2\sin^2\theta\Omega_*^4/3c^3\sim 2.5\times 10^{30}\sin^2\theta
B_{*,6}^2 R_{*,9}^6\Omega_{*,-1}^4 \ {\rm erg/s},
\eeq
where $\mu_{*}=B_*R_*^3$ is the magnetic moment of the dipole stellar field,
$\Omega_{*,-1}=\Omega_*/0.1$ s$^{-1}$, $B_{*,6}$ is the stellar polar surface
field strength in units of $10^6$ G, and $\theta$ is the misalignment angle
between the rotation axis and the magnetic axis. 
For AE Aqr with $\Omega_*=0.19$ s$^{-1}$, we expect 
$L_{em}\sim 4\times 10^{30} B_{*,6}^2$ erg/s or for the observed
upper limit $B_{*,6}\sim 5$, $L_{em}\simle 1\times 10^{32}$ erg/s which is
at least two orders of magnitude lower than the observed spin-down power.

The rapid spin-down has been widely attributed to the propeller action in which
the inflowing material is flung out at a radius $R_x$. This radius is likely
to be beyond the corotation radius and is conventionally understood as the
magnetic truncation radius.
The spin-down power due to the propeller action could be estimated as
\beq
L_{prop}\sim {\dot M}R_x^2(GM_*/R_x^3)\propto R_x^{-1},
\eeq
where $R_x$ is the radius at which the accretion flow is expelled. If the
propeller action occurs near the corotation radius
\beq
L_{prop}\sim {\dot M}(GM_*\Omega_*)^{2/3},
\eeq
and the observed spin-down power requires
\beq
{\dot M}\sim L_{prop}/(GM_*\Omega_*)^{2/3}\sim L_{sd}/(GM_*\Omega_*)^{2/3}
\sim 3\times 10^{17} \ {\rm g/s}.
\eeq
In this case, the expelled material is likely to be flung out at a
characteristic speed of $\sim R_c\Omega_*\sim 2.7\times 10^8$ cm.
The expected high velocity outflow has not been detected in AE Aqr.
If the propeller action occurs at the magnetospheric radius 
$R_{mag}\sim 1.4\times
10^{10} {\dot M}_{16}^{-2/7} M_{*,1}^{-1/7} B_{*,6}^{4/7}$ (cf. eq. 6-2 below), 
$L_{sd}$ is accounted for by the propeller action if 
$L_{sd}\sim GM{\dot M}R_o^{-1}$ or
\beq
{\dot M}\sim 8\times 10^{17} B_{*,6}^{4/9}.
\eeq
In short, the mass accretion rate required for the propeller action to
account for the spin-down power is likely to be considerably higher than
$\sim 10^{17}$ g/s. The observed luminosities indicate that the mass accretion
rate is considerably lower than $\sim 10^{17}$ g/s.
Any magnetized model with the Ghosh-Lamb type (e.g., Frank et al. 1992, Yi 1995
and references therein) would essentially lead to
the same conclusion since the torque achieved in variations of the model is 
essentially limited to the above propeller estimate on the dimensional ground.
We therefore conclude that the observed luminosities and the spin-down power
are incompatible if the spin-down power results in the emission of the
observable radiation.

\section{SPIN-DOWN DUE TO GRAVITATIONAL RADIATION}

We have pointed out that the electromagnetic dipole radiation with
$B_*\le 5\times 10^6$ G and the propeller action with ${\dot M}\le 10^{17}$ g/s
are unable to account for the observed spin-down torque. We propose that
the spin-down power does not transform into the observed electromagnetic 
radiation. We consider the gravitational radiation emission as an 
alternative spin-down mechanism. This mechanism could be an attractive one 
since the resulting 
spin-down power does not need to go into the observable electromagnetic
radiation, which effectively avoids the long standing question of non-detection
of the observed large spin-down power.
The gravitational radiation emission would be a particularly interesting
spin-down mechanism if the spin evolution is highly stable and shows no
signs of disturbances in accretion flows or the stellar magnetospheres.
This appears to be the case in AE Aqr.

Although the white dwarf in AE Aqr spins unusually fast, under normal
circumstances, the gravitational radiation emission requires a rather
high non-zero quadrupole moment or a large eccentricity. If we define the
eccentricity as $\epsilon=[1-(R_2/R_1)^2]/[1+(R_2/R_1)^2]$, where $R_2$ and
$R_1$ are radial extents of the star in the plane perpendicular to the
rotational axis, the gravitational radiation power becomes
\beq
L_{gr}={32G\over 5c^5}\epsilon^2 I_*^2\Omega_*^6.
\eeq
The spin-down power of AE Aqr is accounted for by $L_{gr}$ when
\beq
4\pi^2I_*{\dot P}_*P_*^{-3}={32G\over 5c^5}I_*^2(2\pi/P_*)^6 \epsilon^2,
\eeq
or
\beq
\epsilon=(5c^5{\dot P}_*P^3/32G I_*)^{1/2}\sim 1,
\eeq
which essentially implies that the non-axisymmetric distortion 
of the white dwarf has to
be too large to account for the observed spin-down power.

We have argued that a significant mass accretion must have occurred in order
to account for the observed unusual spin period (e.g. eqs. 2-1, 2-2). 
One of the plausible 
possibilities is that the accreted material mostly lands on a small fraction
of the total surface area near the magnetic poles. If this is the case as
expected in significantly magnetized accretion case, the accreted material would
provide a source of non-zero quadrupole moment if the magnetic axis is 
misaligned with the rotation axis. That is, the magnetically channeled material
would spread from the magnetic poles while its spread is partially hindered 
by the strong stellar magnetic field. 
Conceivably, in a steady state achieved in the
high mass accretion rate episode while the rapid spin-up occurred, prior to
the present spin-down, the accreted material could form accretion mounds
at the magnetic poles (e.g. Inogamov \& Sunyaev 1999 and references therein).

If we assume that the accreted material is present at the magnetic poles in the
form of the spatially limited blobs or mounds while the rotation axis and
the magnetic axis are misaligned by an angle $\theta$, the time averaged
rate of gravitational radiation power could be estimated as
\beq
L_{gr}={336G\over 5c^5}{\delta m}^2 R_*^4 \sin^4\theta\Omega_*^6,
\eeq
where $\delta m$ is the amount of mass accumulated on one magnetic pole.
We have assumed for simplicity that the accumulated material exists at the
magnetic poles without any significant spatial spread and it remains unperturbed
during each stellar rotation. We note that the large coefficient in the
formula effectively amounts to a large eccentricity despite the fact that
the bulk of the stellar mass is not perturbed and distributed in such a way
that the contribution to the quadrupole moment is negligible.

By comparing $L_{gr}$ calculated above and the observed spin-down power
of AE Aqr,
\beq
4\pi I_* {\dot P}_*P_*^{-3}={336G\over 5c^5}{\delta m}^2 R_*^4 \sin^4\theta
\left(2\pi\over P_*\right)^6,
\eeq
we get
\beq
\delta m \sim 1.6\times 10^{-3}\sin^{-2}\theta \ {\rm M_{\odot}},
\eeq
which is roughly $1.6\times 10^{-2}$ of the minimum mass required for the
spin-up of AE Aqr during the rapid accretion phase.

This power becomes rapidly negligible as the star slows down due to the
sensitive dependence of the gravitational radiation power on the spin frequency.
For the AE Aqr parameters, $L_{gr}>L_{em}$ occurs if
\beq
\delta m > 2.1\times 10^{-5} \sin^{-1}\theta B_{*,6} \ {\rm M_{\odot}},
\eeq
or for $\delta m\sim 10^{-3}\ {\rm M_{\odot}}$
\beq
P_* < 0.4 \sin \theta B_{*,6}^{-1} \ {\rm hr}.
\eeq
AE Aqr could well have been continuously spun-down after reaching a high
spin frequency resulting from the high accretion phase. The above estimate
indicates that the AE Aqr's current rapid spin-down could continue to
periods much longer than the present short spin period.

What kinds of magnetized white dwarf systems could show large spin-down 
powers caused by the gravitational radiation emission as in AE Aqr? Obviously
the candidate systems have to be rapidly spinning, which is likely when the
systems were spun-up by preceding high mass accretion rate flows. 
With strong magnetic fields, 
a significant fraction of the accreted mass could reside in the polar
regions while the magnetic axes have to be substantially misaligned with the
rotation axes. For these systems, if the mass accretion becomes high and
the magnetic fields are strong, the propeller-like torque is likely to dominate
over the gravitational radiation emission. The accumulated mass near the 
magnetic poles could spread over the stellar surface although the details
of the spreading process could be highly complicated 
(e.g. Inogamov \& Sunyaev 1999). The gradually
cooling accretion mounds could be a source of persistent UV emission which 
should show significant pulse signals (see below).

\section{ELECTROMAGNETIC EMISSION COMPONENTS AND PHYSICAL PARAMETERS OF
AE AQR}

Based on the large pulse fraction, as high as $\sim$80\% during quiescence
(Eracleous et al. 1994), the most likely UV production site is the 
magnetic poles.
The total UV luminosity (pulsed and non-pulsed) in quiescence is
$L_{uv}\sim 4\times 10^{31}$ erg/s which corresponds to the nominal
polar accretion rate of ${\dot M}\sim 2\times 10^{14}$ g/s.
If the UV emission is the result of the low ${\dot M}$ accretion occurring
at the magnetic poles, the expected emission temperature in the form of the
thermalized radiation is $T_{uv}\sim 2\times 10^{4}a^{-1/4}
({\dot M}/10^{14} {\rm g/s})^{1/4}$ K where $a\le 1$ is the fraction of the 
stellar surface area accreting near the magnetic poles.

On the other hand the X-ray pulse fraction is much lower ($\sim 30$\%) during
quiescence (Choi, Dotani, \& Agrawal 1999).
The straightforward implication from this low pulse fraction is that the 
X-ray emission site is much more extended than the narrow polar region around 
the magnetic poles.

If the accretion occurs in the form of the accretion stream directly impacting
on the magnetosphere, only a small fraction of the accreted material lands
on the magnetic poles. We call such an accretion flow as the high altitude 
accretion flow which constitutes a small fraction of the total accretion flow.
That is, we consider a picture in which a large fraction of the accretion stream
hits low altitudes and a comparable fraction of it hits high altitudes 
and travels directly to the poles (Figure 1).

\subsection{Low Altitude Accretion} 

The power resulting from the low altitude stream-magnetosphere interaction 
region is simply
\beq
L_{x}\sim GM_*{\dot M}_x/R_x,
\eeq
where we denote the location and luminosity by subscript $x$ assuming that
most of the energy is released in the X-ray range and a certain fraction of
the inflowing matter is expelled. X-ray emission is possible
if the accretion stream's kinetic energy is virialized after the accreted
material gets shocked at the impact region. 

The X-ray emission region is characterized by the simple magnetospheric radius
(e.g. eq. 6-2)
\beq
R_{mag}\sim 1.4\times 10^{10}{\dot M}_{16}^{-2/7} M_{*,1}^{-1/7} B_{*,6}^{4/7}
\ {\rm cm},
\eeq
where ${\dot M}_{16}={\dot M}/10^{16}$ g/s. At this radius, the low altitude
stream's ram pressure becomes roughly equal to the magnetic pressure of the 
magnetosphere. We have assumed
that the stream is nearly free-falling and the stream's geometric cross-section
is comparable to the spherical surface area at the interaction region.

If most of the UV emission is due to the low ${\dot M}$ accretion occurring
at high altitudes, we expect as before,
\beq
L_{uv}\sim GM_*{\dot M}_{uv}/R_*.
\eeq
Therefore, we get
\beq
L_{uv}/L_x\sim ({\dot M}_{uv}/{\dot M}_x)(R_x/R_*),
\eeq
or for the magnetospheric radius of the low altitude accretion 
$R_x\sim R_{mag}$.

Using the observed $L_{uv}/L_x\sim 6$ (Eracleous et al. 1994; Choi, Dotani, \&
Agrawal 1999)
\beq
{\dot M}_{uv}/{\dot M}_x\sim 0.3 {\dot M}_{16}^{2/7} M_{*,1}^{-1/7} B_{*,6}^{-4/7},
\eeq
which is uncertain due to the uncertain $B_*$ and ${\dot M}$ in AE Aqr.

If most of the propeller action occurs near the corotation radius $R_c\sim
1.4\times 10^9$ cm, which would be close to the magnetospheric radius $R_{mag}$
if $B_*\sim 2\times 10^4{\dot M}_{16}^{1/2} M_{*,1}^{-1/7}$, then,
the accretion rates in the two regions are compared as
\beq
{\dot M}_{uv}/{\dot M}_x\sim R_*/R_c\sim 0.5,
\eeq
which implies that a large fraction of the accreted matter has to flow to 
the poles.

If the X-ray emission is from the shocked gas at the propeller action region
where the accretion stream hits the magnetosphere and collides with 
the outflowing material creating localized shocks,
the characteristic X-ray 
emission temperature from the optically thin gas is likely to be
\beq
T_x\sim 3GM_*m_p/16kR_x\sim 2\times 10^7 {\dot M}_{16}^{2/7}B_{*,6}^{-4/7} 
\ {\rm K},
\eeq
or $kT_x\sim 2{\dot M}_{16}^{2/7}B_{*,6}^{-4/7}$ keV, which is very close to 
the observed X-ray emission temperature $\simle 3$ keV.
Similarly for the propeller action occurring near the corotation radius,
the expected X-ray emission temperature $kT_x\sim 9$ keV which implies that
the propeller action occurring anywhere between the two regions at radii
$\sim 10^{10}$ cm and $\sim R_c\sim 1.4\times 10^9$ cm can account for the
bremsstrahlung emission in the X-ray band.

On the other hand, if propeller action region is responsible for blackbody
like emission, then the characteristic temperature could be as low as
$\sim 1\times 10^4{\dot M}_{16}^{1/4}$. This temperature corresponds to 
optical/UV emissions although the high pulse fraction observed in AE Aqr
rules out the possibility that the dominant optical/UV emissions arise
from the propeller action region.

In the polar regions, the radially falling material lands on the surface
of the white dwarf and the kinetic energy could thermalize and be radiated
as blackbody like emission. If the accreting fraction of the white dwarf
is $a(\le 1)$ of the total stellar surface area, then the expected emission
temperature is
\beq
T_{uv}\sim 2\times 10^4 a^{-1/4} {\dot M}_{14}^{1/4} \ {\rm K},
\eeq
where ${\dot M}_{14}={\dot M}/10^{14}$ g/s as noted earlier.

Using the observed X-ray luminosity and the X-ray emission temperature,
we can estimate the mass accretion rate and the magnetic field strength.
First, using the shock temperature in the propeller action region,
we require that the shock temperature be close to the observed X-ray
emission temperature $\sim 3$ keV
\beq
{\dot M}_{16}\sim 4.1 B_{*,6}^2.
\eeq
Similarly, the observed X-ray luminosity should be close to the
total accretion power (with the radiative efficiency $\eta_x$) 
$\sim 10^{31}$ erg/s or
\beq
{\dot M}_{16}\sim 0.2 \eta_x^{-7/9}B_{*,6}^{4/9}.
\eeq
The two constraints are simultaneously satisfied only if
$B_*\sim 1.4\times 10^{5}\eta_x^{-1/2}$ G and ${\dot M}\sim 8\times 10^{14}
\eta_x^{-1}$ g/s.

Substituting the mass accretion rate and the field strength in the equation
for the ratio of ${\dot M}_{uv}/{\dot M}_x$ we arrive at
\beq
{\dot M}_{uv}\sim 4\times 10^{14}\eta_x^{-1} \ {\rm g/s},
\eeq
which is a substantial fraction of the accreted material arriving at the
propeller action region. We therefore conclude that a significant fraction
of material manages to land on the surface of the white dwarf despite
an ongoing propeller action in AE Aqr.

\subsection{High Altitude Accretion}

In the above calculations, we have considered that a part of the
accretion stream hits low magnetic altitudes and most of the
X-ray emission is produced from the shock-heated (optically thin)
gas near the magnetospheric radius or the corotation radius. 
On the other hand, the high altitude accretion flow travels 
directly to the poles. At low altitudes, the accreting stream remains mostly
optically thick until shock-heated and expelled by the propeller action.
On the other hand, the accretion stream remains optically thin at high altitudes
and is continuously heated in a roughly quasi-spherical accretion pattern.
The propeller action is relatively weaker at higher altitudes as in the
case of the diskless accretion and hence, a substantial fraction of it 
travels to the poles while emitting X-rays until it reaches the magnetic
poles where the accretion flow gets thermalized and emits optically thick
blackbody-like UV emission. In this picture, because the high altitude gas
stream is adiabatically heated to an X-ray emitting temperature near 
the magnetic poles, we expect pulsed X-ray emission. 

For the derived magnetic field strength, the dipole radiation power becomes
$L_{em}\sim 8\times 10^{28} \eta_x^{-1}$ erg/s which suggests that the
observed radio emission could be accounted for by the dipole radiation
if the X-ray efficiency is $\sim$ 10\%. Alternatively, the radio emission could
result from accelerated electrons at the shock near $R_x$, a remote possibility
given the fact that there exists no evidence of high speed material in AE Aqr.

It is unlikely that the TeV emission occurs at the level of 
$\sim 10^{32}$ erg/s based on the above conclusion. Meintjes et al. (1992, 1994)
and Bowden et al. (1992)
reported detection of TeV $\gamma$-rays from AE Aqr which are pulsating
at the spin period. 
However, according to a more recent observation by Lang et al. (1998), 
there is no evidence for any steady, pulsed or episodic TeV emission.
In the present picture, the TeV emission at the claimed level is not likely
since the largest power goes into the gravitational radiation.

Using the mass accretion rate and the magnetic field strength, we estimate that
the accretion stream is likely to be stopped at a radius
\beq
R_x\sim 1\times 10^{10} \ {\rm cm},
\eeq
which is compared with the nominal circularization radius for material traveling
through the inner Lagrange point in the binary system (e.g. Frank et al. 1992)
\beq
R_{circ}\sim 3\times 10^{10} \ {\rm cm},
\eeq
where we have adopted the secondary to primary mass ratio $\sim 0.6$.
For the estimated magnetospheric radius, it is possible that the accretion
disk doesn't exist in this system as has been recently argued based on the
non-detection of the double-peaked H$\alpha$ emission line in AE Aqr.
Even if they exist, they could exist in a very narrow radial zone in which the
line emission is too weak to detect.

\section{FLARES AND LOW ENERGY EMISSION}

Simultaneous observations between optical and UV and between optical and  
X-ray (Osborne et al. 1995; Eracleous \& Horne 1996) show some similarities
in their light curves.
For example, while UV \& optical flares are well correlated, they are
less correlated with the X-ray flares. However, radio flares are not
correlated with the optical flares (Abada-Simom et al. 1995). 
The pulse amplitudes in both UV and X-ray region do not display any large
variations compared with the values of their quiescent states, nor do they
follow the variations of the non-pulsed level. 
The difference between quiescent and flare spectra in X-ray region is not 
significant, although a hint of spectral hardening is recognized
(Choi, Dotani, \& Agrawal 1999).
These characteristics strongly suggest that the magnetic poles or 
an accretion columns above the white dwarf surface are hardly connected with
the flaring activities.
A possibility that a stellar flare which might occur in the companion 
star has been ruled out because it could not account for the kinematic
properties of line-emitting gas in the UV data (e.g., Eracleous \& Horne 1996).
The flares arising from the occasional encounters between propeller-expelled
outgoing gas streams (or blobs) and the incoming accretion stream are not
realistic because the possible interaction region is too far to
account for the high energies observed in the flares.

The exact nature of the flaring activities is beyond the scope of the present
study. If the flaring activities indeed occur in a region far beyond the polar
area, then a possible region is again the impact region formed by the accretion
stream interacting with the magnetosphere. If the gas rotating
between $R_{circ}$ and $R_x$ wraps the poloidal component, $B_z$,
of the stellar field and amplifies the toroidal component, $B_{\phi}$
to an equilibrium value $B_{\phi}\sim (\gamma/\alpha)(\Omega_*/\Omega-1)B_z$
at a rate $\sim\gamma(\Omega_*-\Omega)B_z$  (e.g. Yi 1995 and reference
therein), it would take $t_{amp}\sim (\alpha\Omega)^{-1}$ where $\alpha\sim
0.1$ is the usual $\alpha$ viscosity parameter, $\Omega_*=2\pi/P_*$,
$\Omega=(GM_*/R^3)^{-1/2}$ is the local rotational angular velocity, and
$\gamma$ is the vertical velocity shear parameter of order unity. Here
we have assumed that the amplification is balanced by the magnetic diffusion
with the magnetic Prandtl number of order unity. Then,
a simple-minded characteristic time scale for magnetic amplification is
$\sim 10\times P_*\sim 300$ s provided that the amplification occurs near the
corotation radius. On the other hand, based on the AE Aqr parameters
${\dot M}\sim 10^{15}$ g/s and $B_*\sim 10^5$ G, we estimate that $R_{mag}\sim
8\times 10^9$ cm and $L_{sd}\sim 1\times 10^{31}$ erg/s. The observed typical
flare energy $\sim 10^{35}$ erg requires the accumulation of the spin-down
energy (in the form of the magnetic stress) for a duration of $\sim 10^4$ s
which is longer than the magnetic field amplification time scale at least
by a factor of $\sim 30$. If the magnetic field amplification is not balanced
by the magnetic diffusion of the $\alpha$ type  (Yi 1995) but by the buoyant 
loss of Parker type (e.g. Wang 1987), 
it is conceivable that the flare energy accumulation
time scale could be accounted for since the field can grow to a higher value
on a longer time scale. The details of this issue is especially hard to 
describe given the uncertain and complex nature of the accretion flow 
pattern and density structure near the propeller action region.

\section{DISCUSSION AND SUMMARY}

Various models have been considered to explain the pulsations and flares 
in AE Aqr. Among the proposed models, both an oblique  rotator model
(Patterson 1979, 1994) and a magnetospheric gating model 
(van Paradijs et al. 1989; Spruit \& Taam 1993) have been ruled out
by the reason that there is lack of observational evidence for 
an accretion disk (see Eracleous \& Horne 1996 and Welsh, Horne \& Gomer
1998 for a detailed discussion).
The propeller model is primarily based on this point.

A high-velocity gas which is escaping from the system is predicted from the
propeller model. However, Welsh, Horne, \& Gomer (1998) have argued that
they did not detect any signatures for the high-velocity gas, nor did
they obtain any expected pattern from their trailed spectrograms.
Therefore, without any evidence for the high-velocity gas, it is rather
unlikely that the propeller action is mainly responsible for the
spin-down of AE Aqr. 
We also note that in the propeller scenario mass-loss from the companion
star is inhomogeneous and intermittent. If this is true, it is
questionable why the pulsed emission is sustained almost constantly and
stably.

By removing the spin-down power from the observable luminosities, we have
constructed a self-consistent model which in essence incorporates all the
existing ingredients.
Eracleous et al. (1994) analyzed 10 pulse profiles in different wavelengths
obtained from the simultaneous HST (UV) and ground (optical) observations. 
According to their results, there is no discernible
phase shifts between pulses at different wavelengths, and the amplitude of
pulsations decreases with increasing wavelength.
This implies that the emission region becomes broader for longer wavelengths
(or lower energies). 
In our model, pulsed X-ray emission is from the adiabatically heated gas in 
an extended region between the high altitude magnetosphere and the magnetic 
poles, which is traveling onto the poles.
While the gas travels toward the poles, it cools down due to the
radiation and UV emission becomes dominant near the polar region.
After the gas landed on the polar regions, it would spread over the
stellar surface as discussed in \S\ 3.
While spreading over the stellar surface, the gas cools down further.
Therefore, it is possible that the emission region becomes broader for
longer wavelengths.
On the other hand, the incoming accreting stream is optically thick at low
altitudes and it is mostly expelled. Therefore, optical/UV emission
can also arise in this region and partly contribute to the non-pulsed
optical/UV emission.
Alternatively, if cooling occurs rapidly during the gas infall phase,
the kinetic energy of the radially falling material could thermalize
at the polar region and be radiated as blackbody like emission with
temperature of $\sim 2\times10^4$ K as estimated in eq. 4-8.
The gas then follows the similar spreading and cooling processes on the
white dwarf surface.

We have considered a possibility that the flares of AE Aqr are due to the 
release of magnetic stress which is reminiscent of the solar flares.
An alternative possibility, the magnetic pumping model, was proposed by 
Kuijpers et al. (1997) with an intention to account for radio outbursts
or flares.
In this model, the radio flares are caused by eruptions of bubbles of
fast particles from a magnetosphere surrounding the white dwarf.
The model also speculates that at relatively low accretion rates
the conversion of spin energy into acceleration (rather than heating)
of electrons and protons can be efficient. The accelerated fast particles
remain trapped in the magnetosphere and when their total energy
becomes comparable to the magnetic field energy, an MHD instability sets in.
Then the synchrotron radiation occurs in the expanding plasmoid at radio and
long wavelengths.

The highly stable spin-down implies that the spin-down mechanism
remains stable despite propeller action and related possible dynamical 
instabilities.
It has usually been interpreted as a sign of a stable accretion disk. 
If AE Aqr doesn't have an accretion disk as often claimed, 
the spin-down could indeed be
due to the gravitational radiation emission which is obviously quite stable
as long as it is the dominating spin-down mechanism.

If there exists a Keplerian accretion disk interacting with the stellar
magnetosphere, the accretion flow will exert a torque (Yi 1995)
\beq
N={7N_o\over 6}{1-(8/7)(R_{mag}/R_c)^{3/2}\over 1-(R_{mag}/R_c)^{3/2}},
\eeq
where $N_o={\dot M}(GM_*R_{mag})^{1/2}$ and $R_{mag}$ is the magnetic
truncation or the magnetospheric radius determined by
\beq
(R_{mag}/R_c)^{7/2}=A\left|1-(R_{mag}/R_c)^{3/2}\right|,
\eeq
where
\beq
A=2(\gamma/\alpha)B_c^2R_c^3/{\dot M}(GM_*R_c)^{1/2},
\eeq
with $B_c=\mu_*/R_c^3$. The observed spin-down will result if
$R_{mag}/R_c>0.9148$ in this particular magnetized accretion model 
or $A>5.86$ which for AE Aqr translates into
\beq
B_*>1\times 10^4(\alpha/\gamma)^{1/2}{\dot M}_{15}^{1/2}
\eeq
where ${\dot M}_{15}={\dot M}/10^{15}g/s$. Therefore, for the derived AE Aqr
parameters, a rotating accretion disk would always exert a spin-down torque
although the spin-down is dominated by the gravitational radiation mechanism.

Our proposed model is essentially summarized as follows. (i) The spin-down is
driven by gravitational radiation emission. (ii) The dipole radiation from the
rapidly spinning white dwarf is responsible for radio emission and possible 
TeV $\gamma-$ray emission.
(iii) UV emission is from the magnetic poles with a low accretion rate or 
from cooling accretion mound from previous high accretion episode. 
A large amount of material
present at the poles would be compatible with the current rapid white dwarf 
rotation.
(iv) Accretion to the poles occur at high altitudes while at low altitudes
the propeller action drives material outward. 
However, as long as the accretion rate
is lower than $\sim 10^{17} g/s$, the spin-down power by propeller action is
smaller than the observed spin-down. 
(v) X-rays are mostly from the accretion stream-magnetosphere boundary where 
the propeller action drives material outward while shock-heating the accreted
material. UV emission could also arise in this region if there exists optically
thick gas.
(vi) Flares are due to the release of magnetic stress much like that seen in 
solar flares.
(vii) In sum, $L_{sd}\sim 2\times 10^{34}$ erg/s mostly goes to $L_{gw}$ which
is undetectable. $L_{em}\simle 10^{32}$ erg/s goes to $L_{radio}$ and possibly
to $L_{\gamma}$. $L_{uv}$ is either from ${\dot M}\sim 2\times 10^{14}$ g/s high
altitude accretion or from the cooling accretion mound at the magnetic poles.
The contribution from the disk-magnetosphere interaction region has to be
small as required by the high pulse fraction. 

\acknowledgments

IY is grateful to John Bahcall and Craig Wheeler for hospitality 
and Bill Welsh for useful information. IY thanks Korea Research
Foundation for its grant support KRF 1998-001-D00365 for its generous support. 
CSC is grateful to the Korea Astronomy Observatory for its grant support 
KAO 99-1-201-20.


\clearpage
\noindent
\begin{center}
{\bf FIGURE CAPTIONS}
\end{center}

\figcaption[]{A schematic description of the AE Aqr accretion-ejection model
as discussed in the text.
}

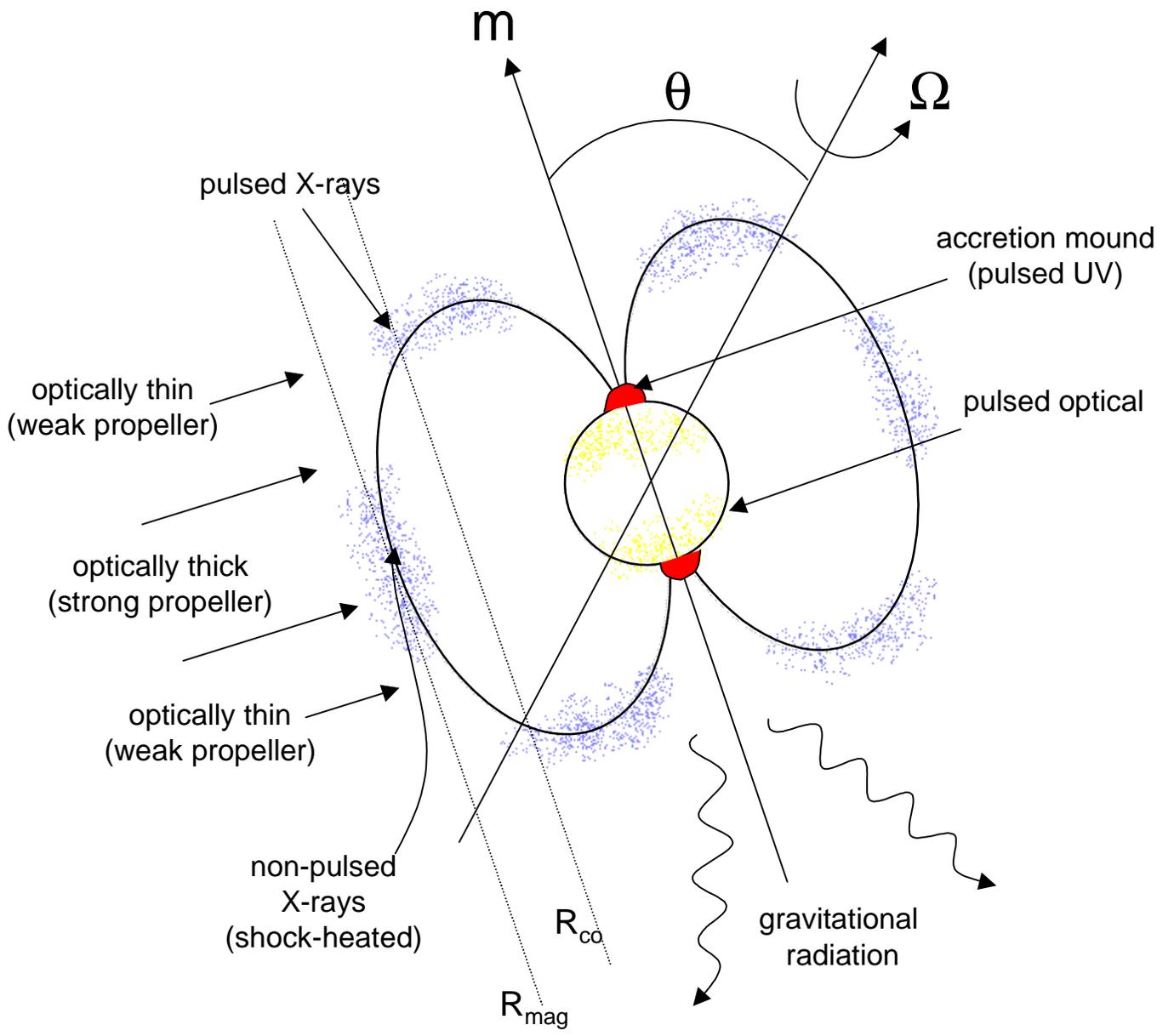
\end{document}